\documentclass[10pt,aps,prl,twocolumn,superscriptaddress,floatfix,longbibliography, groupedaddress]{revtex4-1}

\usepackage{graphicx}
\usepackage{amsfonts}
\usepackage{amssymb}
\usepackage{amsmath}
\usepackage{txfonts}
\usepackage{lipsum}
\usepackage{color}
\usepackage{wasysym}
\usepackage[colorlinks=true, allcolors=blue]{hyperref}
\usepackage{bbold}
\usepackage{etoolbox} 
\usepackage[normalem]{ulem}
\usepackage{mathtools} 
\usepackage{multirow} 
\usepackage{hhline}
\usepackage{mathrsfs} 
\usepackage{soul}

\usepackage{letltxmacro}
\LetLtxMacro{\oldsqrt}{\sqrt}
\renewcommand{\sqrt}[2][\mkern8mu]{\mkern-6mu\mathop{}\oldsqrt[#1]{#2}}

\usepackage{url}
\definecolor{indigo(dye)}{rgb}{0.0, 0.25, 0.42}
\usepackage{placeins}

\begin{document}

\title{Eliminating Orbital Selectivity from the Metal-Insulator\\
Transition by Strong Magnetic Fluctuations}

\author{Evgeny~A.~Stepanov}
\email{evgeny.stepanov@polytechnique.edu}
\affiliation{CPHT, CNRS, Ecole Polytechnique, Institut Polytechnique de Paris, F-91128 Palaiseau, France}

\begin{abstract}
The orbital-selective electronic behavior is one of the most remarkable manifestations of strong electronic correlations in multi-orbital systems.
A prominent example is the orbital-selective Mott transition (OSMT), which is characterized by the coexistence of localized electrons in some orbitals, and itinerant electrons in other orbitals.
The {\it state-of-the-art} theoretical description of the OSMT in two- and three-dimensional systems is based on local non-perturbative approximations to electronic correlations provided by dynamical mean-field theory or slave spin method.
In this work we go beyond this local picture and focus on the effect of spatial collective electronic fluctuations on the OSMT.
To this aim, we consider a half-filled Hubbard-Kanamori model on a cubic lattice with two orbitals that have different bandwidths.
We show that strong magnetic fluctuations that are inherent in this system prevent the OSMT and favor the N\'eel transition that occurs at the same critical temperature for both orbitals.
\end{abstract}

\maketitle

Materials that are characterized by strong interactions between electrons often demonstrate a non-trivial multi-orbital character of electronic correlations.
For instance, some of these materials display signatures of the orbital-selective Mott transition (OSMT), where itinerant and localized electrons that live in different orbitals coexist with each other~\cite{GOODENOUGH1982383, anisimov2002orbital, PhysRevLett.101.126401, Kou_2009, de2009genesis, Hackl_2009}.
The orbital-selective character of the paring mechanism is also considered to be one of the most important ingredients for the formation of the superconducting state in ruthenates~\cite{PhysRevLett.78.3374, RevModPhys.75.657, mackenzie2017even} and iron-based superconductors~\cite{PhysRevLett.101.087004, PhysRevB.79.014505, PhysRevLett.111.107006, yin2014spin, baek2015orbital, PhysRevLett.116.247001, PhysRevB.95.174504, doi:10.1126/science.aal1575}.
An accurate description of these effects can be performed only on the basis of advanced theoretical methods.
For instance, addressing the Mott transition requires a non-perturbative treatment of local electronic correlations, which can be carried out in the framework of the dynamical mean-field theory (DMFT)~\cite{RevModPhys.68.13, RevModPhys.78.865} or the slave-spin~\cite{PhysRevB.72.205124, PhysRevB.81.035106, PhysRevB.86.085104} approaches.
Using these methods in the case of realistic two- and three-dimensional multi-orbital systems is already expensive numerically and represents the current {\it state-of-the-art} description of the OSMT~\cite{PhysRevB.79.115119, PhysRevLett.102.126401, PhysRevB.84.195130, PhysRevB.86.035150, Wang_2016, PhysRevB.100.115159}.

DMFT and slave-spin methods are usually successful in describing physical effects related to local electronic correlations, but neglect spatial collective electronic fluctuations.
Insights from model single-band systems, that are explored more significantly~\cite{RevModPhys.90.025003, PhysRevX.11.011058}, suggest that non-local correlations may noticeably affect the physics of the system.
For instance, strong magnetic fluctuations significantly reduce the critical value of the Coulomb interaction for the metal-insulator transition compared to DMFT~\cite{PhysRevLett.101.186403, PhysRevB.98.155117, PhysRevB.100.205115}.
In multi-orbital systems, taking into account the non-local collective electronic fluctuations may result in the redistribution of the spectral weight between different orbitals, which completely changes the scenario for the metal-insulator transition predicted by DMFT~\cite{2022arXiv220402116V}.
In addition, in realistic systems the OSMT is often accompanied by the presence of magnetic fluctuations~\cite{PhysRevLett.104.057002, yin2011kinetic, PhysRevLett.116.247001, benfatto2018nematic, herbrych2018spin, patel2019fingerprints, PhysRevB.102.054430, PhysRevB.104.125122, PhysRevLett.127.077204, PhysRevB.105.075119, bai2022antiferromagnetic}. 
Moreover, in three dimensions the Mott insulating phase is usually hidden inside the antiferromagnetic (AFM) state, which also does not allow one to disregard magnetic fluctuations when studying the OSMT.

In the case of one-dimensional systems, the density matrix renormalization group method allows one to study the OSMT in the presence of the non-local correlations and, in particular, of a magnetic state~\cite{Plekhanov_2011, PhysRevLett.112.106405, PhysRevB.94.235126, PhysRevB.96.155111, herbrych2018spin, patel2019fingerprints, PhysRevB.104.125122, PhysRevLett.127.077204, PhysRevB.105.075119}.
However, in higher-dimensions, where the physics can be substantially different from a one-dimensional case, the effect of the non-local collective electronic fluctuations on the OSMT remains poorly understood.
Indeed, describing two- and three-dimensional multi-orbital systems beyond local approximations using accurate diagrammatic techniques remains extremely expensive from the computational point of view~\cite{PhysRevLett.107.137007, PhysRevLett.113.266403, PhysRevB.95.115107, Boehnke_2018, acharya2019evening, PhysRevB.100.125120}.
On the other hand, relatively simple (non-diagrammatic) extended DMFT approaches do not capture the non-local nature of magnetic fluctuations originating from the spatial two-particle electronic excitations~\cite{PhysRevB.52.10295, PhysRevLett.77.3391, PhysRevB.61.5184, PhysRevLett.84.3678, PhysRevB.63.115110, PhysRevB.66.085120, napitu2012effects}.
Cluster extensions of DMFT~\cite{PhysRevB.58.R7475, PhysRevB.62.R9283, PhysRevLett.87.186401, RevModPhys.77.1027, doi:10.1063/1.2199446} are also not very suitable for this purpose, because in the multi-orbital case they are able to account for only short-range correlation effects~\cite{PhysRevLett.101.256404, PhysRevB.89.195146, PhysRevB.91.235107}.

In this work, the non-local collective electronic fluctuations in a three-dimensional system are considered in the framework of the dual triply irreducible local expansion \mbox{D-TRILEX} method~\cite{PhysRevB.100.205115, 2022arXiv220406426V}.
This approach represents a relatively inexpensive diagrammatic extension of DMFT~\cite{RevModPhys.90.025003}, where leading channels of instability are considered simultaneously without any limitation on the range.
Single- and two-particle fluctuations in the \mbox{D-TRILEX} theory are coupled by means of the lowest-order three-point vertex corrections that are crucially important for a correct description of the orbital and spatial structure of collective electronic effects~\cite{PhysRevLett.127.207205}.
In this way, the \mbox{D-TRILEX} approach provides a self-consistent consideration of the feedback of collective electronic fluctuations onto single-particle quantities~\cite{PhysRevB.103.245123, stepanov2021coexisting, 2022arXiv220402116V}.

We apply the \mbox{D-TRILEX} method to a two-orbital Hubbard-Kanamori model on a cubic lattice, where the OSMT was previously addressed in the framework of local theories.
We find that in the considered system significant magnetic fluctuations develop already at a relatively high temperature, but in this regime their strength is orbital-dependent. 
Upon lowering the temperature contributions of different orbitals to the spin susceptibility mix more efficiently.
Consequently, the strength of the magnetic fluctuations increases and becomes orbital-independent close to N\'eel temperature.
We find that the N\'eel transition to the ordered AFM phase occurs before the system experiences the OSMT.
Importantly, at the transition point all orbital components of the spin susceptibility diverge simultaneously, which indicates that the N\'eel transition does not display an orbital-selective character.
Moreover, this N\'eel transition also persist in the presence of a non-zero interorbital hopping that is responsible for destroying the OSMT at low temperatures in favor of the metallic ground state~\cite{kugler2021orbitalselective}.

\begin{figure}[t!]
\includegraphics[width=1.0\linewidth]{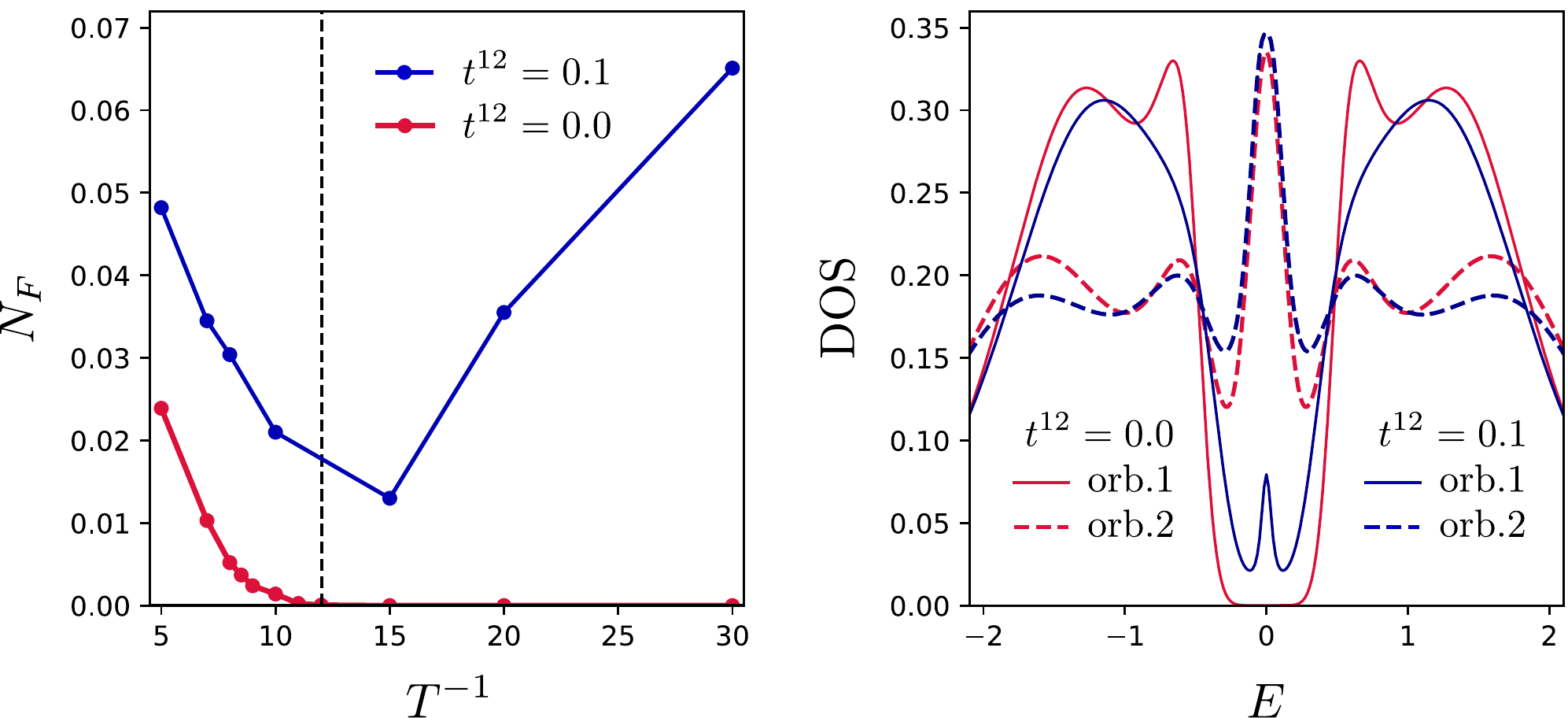}
\caption{Local electronic DOS of DMFT obtained in the absence (red line) and in the presence (blue line) of the interorbital hopping $t^{12}$. Left panel shows the electronic density at Fermi energy $N_F$ computed for the narrow orbital (${l=1}$) as the function of the inverse temperature $T^{-1}$. Vertical dashed line at ${T^{-1}=12}$ indicates the OSMT point. Right panel shows the orbitally-resolved DOS calculated at ${T^{-1}=40}$. Solid and dashed lines correspond to narrow (${l=1}$) and wide (${l=2}$) orbitals, respectively.  
\label{fig:DMFT}}
\end{figure}

{\it Local approximation.} 
A minimal model that allows one to address the OSMT is a half-filled Hubbard model with two orbitals that have different bandwidths.
The Hamiltonian for this model reads
\begin{align*}
H = \sum_{jj',ll',\sigma} t^{ll'}_{jj'} c^{\dagger}_{jl\sigma} c^{\phantom{\dagger}}_{j'l'\sigma} + \frac12 \hspace{-0.05cm} \sum_{j,\{l\},\sigma\sigma'} U_{l_1l_2l_3l_4} c^{\dagger}_{j l_1 \sigma} c^{\phantom{\dagger}}_{j l_2 \sigma} c^{\dagger}_{j l_4 \sigma'} c^{\phantom{\dagger}}_{j l_3 \sigma'}
\end{align*}
For the convenience of describing magnetic fluctuations the interacting part of the Hamiltonian is written in the particle-hole representation. 
The local interaction matrix $U_{l_1l_2l_3l_4}$ is parametrized in the Kanamori form~\cite{10.1143/PTP.30.275, doi:10.1146/annurev-conmatphys-020911-125045} with the intraorbital ${U=U_{llll}}$ and the interorbital ${U'=U_{lll'l'}}$ Coulomb interactions, and the Hund's coupling ${J=U_{ll'll'}=U_{ll'l'l}}$.
Operators $c^{(\dagger)}_{jl\sigma}$ describe annihilation (creation) of an electron on the site $j$, at the orbital $l\in\{1, 2\}$, and with the spin projection $\sigma\in\{\uparrow, \downarrow \}$. 
To be consistent with the previous studies, we consider a simple cubic lattice with nearest-neighbor hoppings $t^{ll'}_{\langle jj' \rangle}$ in the form of Ref.~\onlinecite{kugler2021orbitalselective}. 
The momentum-space representation of the intraorbital $t^{ll}$ and the interorbital $t^{12}$ components of the hopping matrix are
\begin{align*}
t^{ll}_{\bf k}~ = &- 2t^{ll} \left( \cos{}k_x + \cos{}k_y + \cos{}k_z \right) \\
t^{12}_{\bf k} = &- 2t^{12} \left( \cos{}k_x - \cos{}k_y \right)
\end{align*}
The half-bandwidth of the first orbital ${D_1 = 6t^{11} = 1}$ defines the energy scale of the system. 
The second orbital is chosen to be twice wider than the first one, i.e. ${D_2 = 6t^{22}=2}$.
We also stick to the rotationally invariant form of the interaction with ${U=2.4}$, ${U'=U-2J=1.6}$, and ${J=0.4}$ parameters~\cite{kugler2021orbitalselective}.  

\begin{figure}[t!]
\includegraphics[width=0.9\linewidth]{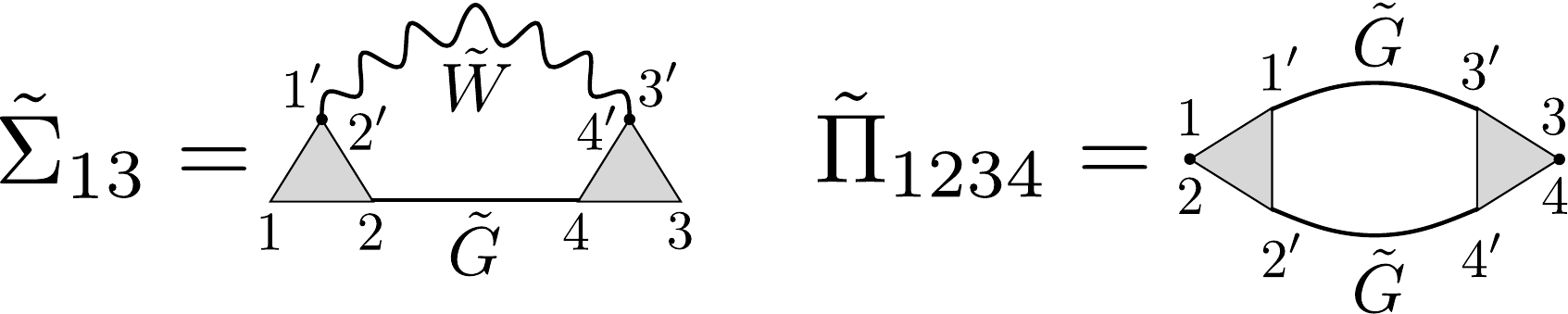}
\caption{Diagrams for the self-energy $\tilde{\Sigma}$ and the polarization operator $\tilde{\Pi}$ of the \mbox{D-TRILEX} approach. 
Orbital dependence is explicitly shown by numbers. 
$\tilde{G}$ and $\tilde{W}$ are respectively the Green's function and the renormalized interaction used in this theory. Gray triangles depict local vertex corrections.
\label{fig:Sigma_Pi}}
\end{figure}

In the absence of the interorbital hopping $t^{12}$ the OSMT has been explored in a large space of model parameters using DMFT and slave spin methods~\cite{PhysRevB.79.115119, PhysRevLett.102.126401, PhysRevB.84.195130, PhysRevB.86.035150, Wang_2016, PhysRevB.100.115159}. 
In the considered two-orbital model the OSMT can be explained as follows.
The presence of the strong Hund’s coupling $J$ suppresses orbital fluctuations in the system, and the correlation effects become strongly dependent on the individual characteristics of each band~\cite{doi:10.1146/annurev-conmatphys-020911-125045, PhysRevB.83.205112}.
In this case, the relative strength of the electronic interaction can be determined by the ratio between the interaction and the width of the corresponding band, which favours the Mott transition for a more narrow orbital.
Remarkably, including the interorbital hopping between the metallic and the insulating orbitals destroys the OSMT and results in a metallic ground state of the system.
This fact was recently reported in Ref.~\onlinecite{kugler2021orbitalselective} on the basis of low temperature DMFT calculations.
We confirm this result by making finite-temperature DMFT calculations using the program package w2dynamics~\cite{WALLERBERGER2019388}.
To obtain the electronic density of states (DOS), we perform the analytical continuation for the local Green's function from Matsubara frequency space to real energies based on the maximum entropy method implemented in the ana\_cont package~\cite{kaufmann2021anacont}.
Left panel in Fig.~\ref{fig:DMFT} shows the corresponding result for the electronic density $N_F$ at Fermi energy ${E=0}$ for the narrow orbital (${l=1}$) as the function of the inverse temperature ${T^{-1}}$.
We find that in the absence of the interorbital hopping (red line) $N_F$ becomes zero around ${T^{-1}=12}$ (vertical dashed black line), and the considered two-orbital system undergoes the OSMT. 
As illustrated in the right panel of Fig.~\ref{fig:DMFT}, below the transition temperature the narrow orbital becomes insulating (solid red line), while the wide orbital (${l=2}$) remains metallic with a pronounced peak in the DOS at ${E=0}$ (dashed red line).
Taking into account ${t^{12}=0.1}$ drastically changes this physical picture. 
In this case, the electronic density $N_{F}$ for the narrow orbital (blue line in left panel of Fig.~\ref{fig:DMFT}) first decreases with decreasing temperature, but after some critical value of $T$ starts to increase again.
As the result, both orbitals are metallic at low temperatures and exhibit a quasiparticle peak at Fermi energy.
These peaks are clearly visible in the orbital-resolved electronic DOS, which for ${T^{-1}=40}$ is plotted in solid (${l=1}$) and dashed (${l=2}$) blue lines in the right panel of Fig.~\ref{fig:DMFT}.
This observation raises two important questions: does the orbital-selective metal-insulator transition exist in the case of a nonzero interorbital hopping that is often present in realistic systems, and are local approximations sufficient for describing this transition?

{\it Effect of non-local magnetic fluctuations.}
To go beyond the local picture, we additionally consider the effect of non-local collective electronic fluctuations using the \mbox{D-TRILEX} approach~\cite{PhysRevB.100.205115, 2022arXiv220406426V}.
This method allows for a self-consistent and simultaneous description of spatial charge and spin fluctuations that are treated diagrammatically beyond DMFT~\cite{PhysRevB.103.245123, stepanov2021coexisting, 2022arXiv220402116V}.
To avoid the double-counting of local correlation effects that are already taken into account in DMFT, the self-energy $\tilde{\Sigma}$ and the polarization operator $\tilde{\Pi}$ in the \mbox{D-TRILEX} theory are constructed in an effective ``dual'' space in terms of corresponding interacting Green's functions $\tilde{G}_{ll'}$, renormalized charge and spin interactions $\tilde{W}^{ch/sp}_{l_1l_2l_3l_4}$, and exact local three-point vertex corrections. 
Diagrammatic expressions for $\tilde{\Sigma}$ and $\tilde{\Pi}$ are shown in Fig.~\ref{fig:Sigma_Pi}.
The dual quantities are calculated self-consistently using the standard Dyson equations.
After that, the physical Green's function $G_{ll'}$ and the charge and spin susceptibilities $X^{ch/sp}_{l_1l_2l_3l_4}$ are obtained using the exact relations between the corresponding dual and lattice quantities~\cite{PhysRevB.100.205115, 2022arXiv220406426V}. 
Multi-orbital \mbox{D-TRILEX} calculations are performed on the basis of the numerical implementation described in Ref.~\onlinecite{2022arXiv220406426V}.

\begin{figure}[t!]
\includegraphics[width=1.0\linewidth]{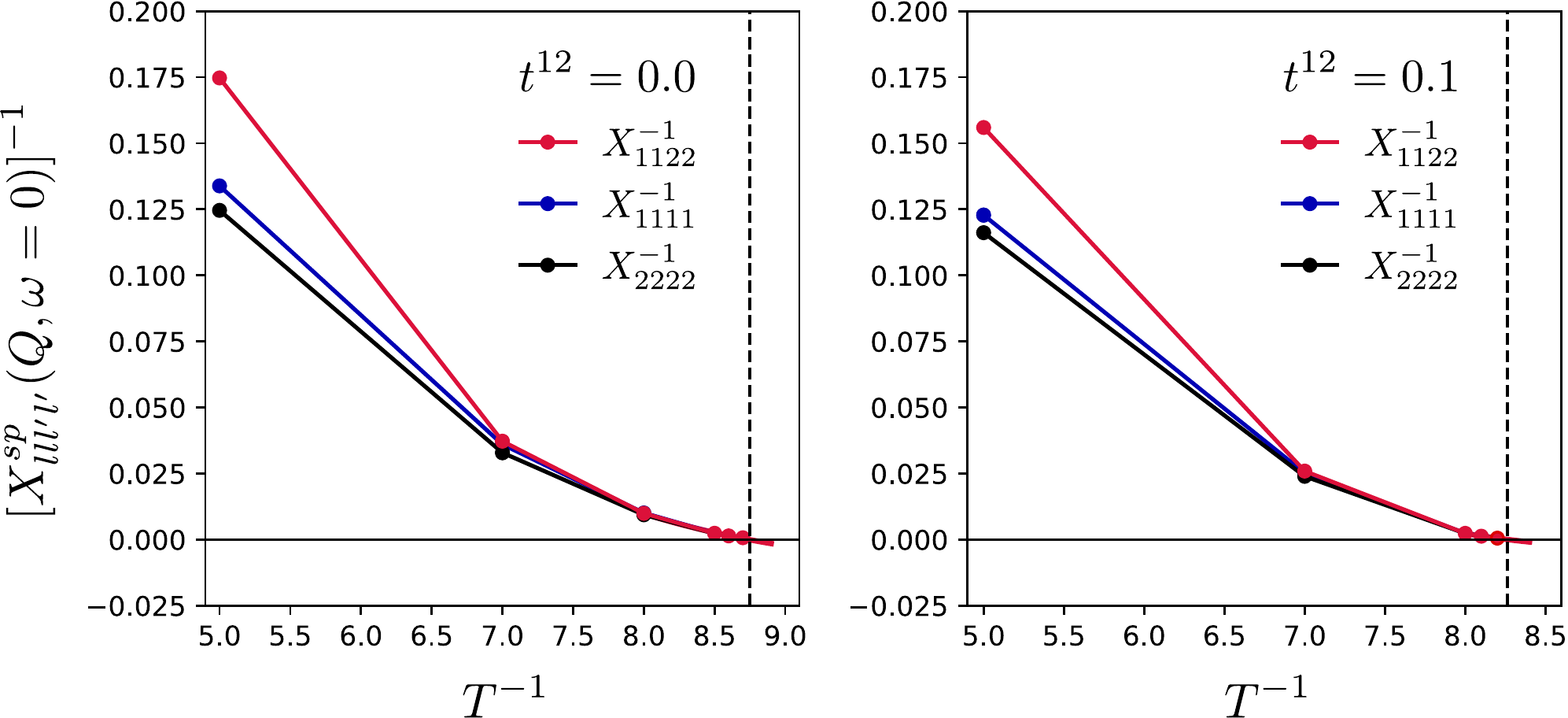}
\caption{Temperature dependence of the intraorbital $X_{1111}$ and $X_{2222}$ and the interorbital $X_{1122}$ components of the spin susceptibility $X^{sp}_{lll'l'}$ calculated at zero frequency $\omega=0$ at the antiferromagnetic wave vector ${Q = \{\pi,\pi,\pi\}}$. Results are obtained for the interorbital hopping ${t^{12}=0.0}$ (left panel) and ${t^{12}=0.1}$ (right panel). The vertical dashed line at ${T^{-1}=8.75}$ (left panel) and  ${T^{-1}=8.25}$ (right panel) indicates the N\'eel transition point at which all the components of the spin susceptibility diverge.
\label{fig:X1234}}
\end{figure}

The momentum-resolved spin susceptibility $X^{sp}_{lll'l'}({\bf q},\omega)$ calculated at zero frequency ${\omega=0}$ displays a strong instability with the maximum value at the wave vector ${Q = \{\pi,\pi,\pi\}}$.
This fact indicates that the considered system exhibits strong AFM fluctuations.
In turn, charge ($X^{ch}_{lll'l'}$) and orbital (${X^{ch}_{ll'll'}=X^{sp}_{ll'll'}}$) fluctuations are weak, and their effect on the OSMT can be neglected.
In a single-orbital case, the critical temperature for the N\'eel transition to the AFM state depends on the relative strength of the interaction compared to the bandwidth (see e.g. Ref.~\onlinecite{PhysRevB.92.144409}). 
For this reason, one can naively expect that in the system, where the orbitals have different bandwidths, strong magnetic fluctuations may result in the orbital-selective N\'eel transition.
However, in practice this can happen only for the totally uncoupled orbitals. 
In a more realistic situation, the susceptibility $X^{ch/sp}_{l_1l_2l_3l_4}$ has a complex orbital structure, in particular, due to the presence of interorbital components $U'$ and $J$ in the interaction matrix. 
When collective electronic fluctuations are strong, contributions of different orbitals to the susceptibility can mix so efficiently that it may completely change the physical behavior of the system~\cite{PhysRevLett.127.207205}.
This fact also applies to the current problem.
Fig.~\ref{fig:X1234} shows the temperature dependence of different orbital components of the AFM susceptibility ${X^{sp}_{lll'l'}(Q,\omega=0)}$. 
For illustrative purposes, we plot the inverse value of each component as a function of the inverse temperature. 
We find that at a relatively high temperature (${T^{-1}=5}$) all three orbital components of the susceptibility are different, with a dominant contribution of the intraorbital $X^{sp}_{llll}$ parts. 
At this temperature the AFM fluctuations are yet not very strong, which is confirmed by a not very large leading eigenvalue ($\lambda$) of the Bethe–Salpeter equation in the orbital space calculated for the spin channel: ${\lambda=0.58}$ for ${t^{12}=0.0}$ and ${\lambda=0.62}$ for ${t^{12}=0.1}$.
Lowering the temperature significantly increases the strength of the AFM fluctuations.
At the inverse temperature ${T^{-1}=8.5}$ for the case of ${t^{12}=0.0}$ and at the ${T^{-1}=8.0}$ for ${t^{12}=0.1}$ the leading eigenvalue of the AFM fluctuations is already ${\lambda=0.99}$.
This fact indicates that the contributions of different orbitals to the susceptibility are thoroughly mixed, which explains the nearly identical values of the $X^{sp}_{1111}$, $X^{sp}_{1122}$, and $X^{sp}_{2222}$ components at low temperatures.
Remarkably, the interorbital hopping does not play an important role in mixing different orbital components in the susceptibility, because this effect takes place even in the absence of $t^{12}$. 
As has been shown in Ref.~\onlinecite{PhysRevLett.127.207205}, the mixing of orbital degrees of freedom mainly comes from the three-point vertex corrections that depend on four orbital indices as explicitly shown in Fig.~\ref{fig:Sigma_Pi}.
Without these vertices, the orbital structure of the polarization operator is determined only by the Green's functions $\tilde{G}$ that in the absence of the interorbital hopping are diagonal in the orbital space.
In this case, the intra- and interorbital contributions to the susceptibility are connected only via the interaction matrix, which does not result in a very efficient mixing of the orbital degrees of freedom~\cite{PhysRevLett.127.207205}.

Fitting the temperature dependence of the AFM susceptibility by an exponential function shows that all three components of the susceptibility diverge at the same temperature.
Indeed, the inverse of each component becomes zero at ${T^{-1}=8.75}$ for the case of ${t^{12}=0.0}$ and at ${T^{-1}=8.25}$ for ${t^{12}=0.1}$. 
This result means that the N\'eel transition for both orbitals occurs simultaneously and does not display an orbital-selective character. 
This observation suggests that the Hund's coupling considered in the model does not decouple the orbital degrees of freedom in the non-local collective electronic fluctuations contrary to what has been found for the local correlation effects~\cite{doi:10.1146/annurev-conmatphys-020911-125045, PhysRevB.83.205112}.  
Remarkably, the critical temperature (${T^{-1}=12}$) for the OSMT in the absence of $t^{12}$ is lower than the N\'eel temperature.
Thus, strong magnetic fluctuations in the considered system prevent the orbital-selective metal-insulator transition and drive the system towards the AFM ordered phase with two insulating orbitals. 
The presence of the interorbital hopping does not change this physical picture.
In this case, the N\'eel transition is shifted to a higher critical temperature, and instead of the metallic state predicted by DMFT~\cite{kugler2021orbitalselective} the system also becomes the AFM insulator.

\begin{figure}[t!]
\includegraphics[width=0.95\linewidth]{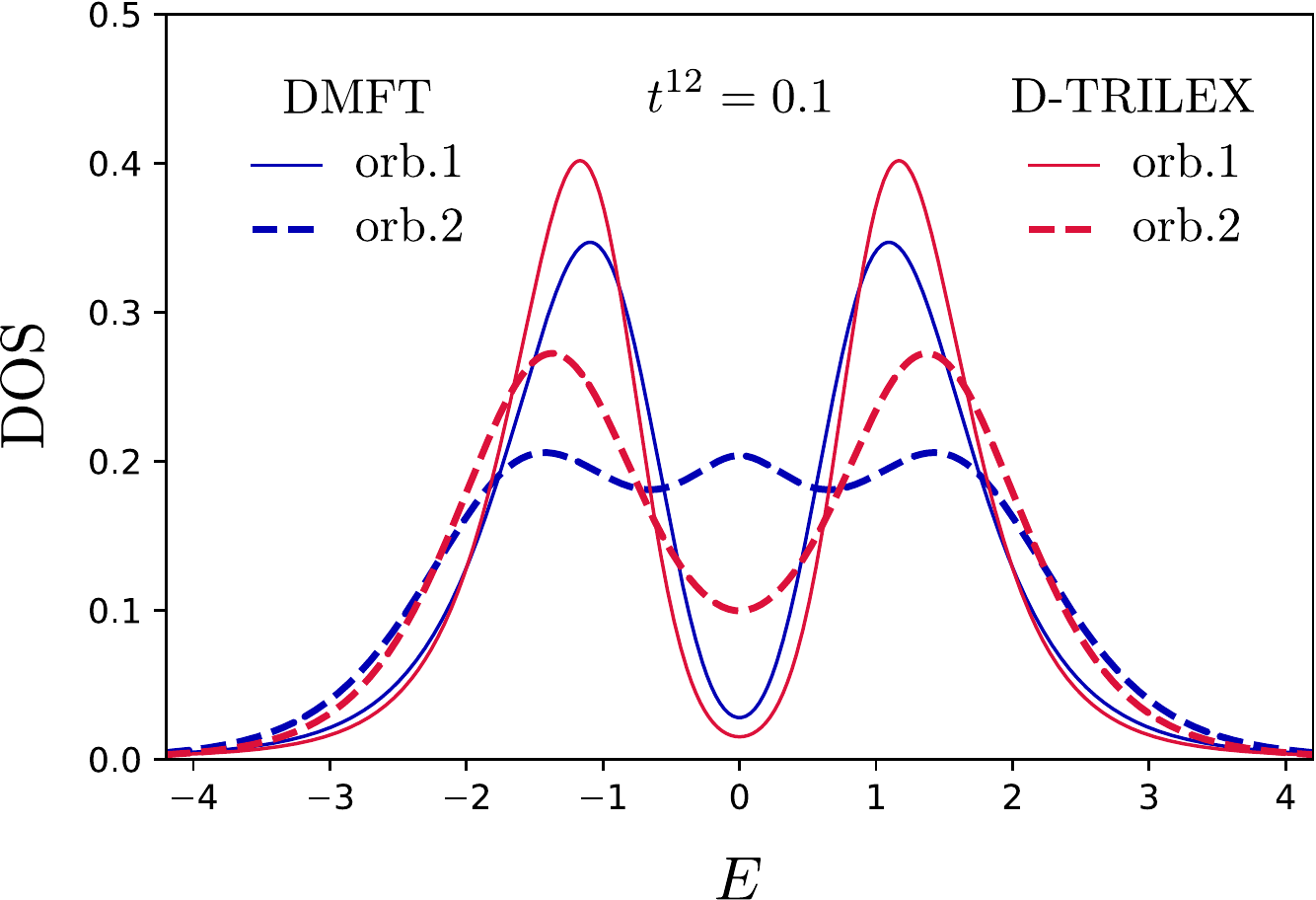}
\caption{The orbitally-resolved local electronic DOS calculated using DMFT (blue lines) and \mbox{D-TRILEX} (red lines) methods. Solid lines correspond to the first (narrow) orbital. Dashed lines depict the DOS of the second (wide) orbital. The results are obtained for the case of ${t^{12}=0.1}$ at the inverse temperature ${T^{-1}=8.2}$ close to the N\'eel transition point.
\label{fig:DOS}}
\end{figure}

The signature of the metal-insulator transition for both orbitals can also be found in the electronic spectral function.
Indeed, the divergence of the susceptibility at the momentum $Q$ that defines the ordering vector in the instability channel results in the divergence of the corresponding renormalized interaction $\tilde{W}$.
This divergent interaction enters the electronic self-energy $\tilde{\Sigma}$ (see Fig.~\ref{fig:Sigma_Pi}) and leads to a formation of a pseudogap in the electronic spectral function~\cite{keldysh1965possible, kozlov1965metal, PhysRevB.97.085125, PhysRevLett.120.216401, stepanov2021coexisting}.
Note that our calculations are performed in the paramagnetic phase at temperatures above the transition point. 
For this reason, we can only capture the formation of the pseudogap, which below the N\'eel temperature becomes a true gap due to the long-range AFM ordering.
Fig.~\ref{fig:DOS} shows the orbital-resolved local electronic DOS obtained for the case of ${t^{12}=0.1}$ at the inverse temperature ${T^{-1}=8.2}$.
This is the closest point to the N\'eel transition where we could converge numerical calculations~\footnote{ 
In the absence of the interorbital hopping the result for the DOS is qualitatively the same and is not shown in this work.}.
DMFT (blue lines) predicts different behavior for the two orbitals.
The narrow orbital depicted by a solid line is almost in the Mott-insulating regime, but has a finite electronic density at Fermi energy according to the result of Ref.~\onlinecite{kugler2021orbitalselective}.
The wide orbital demonstrates a correlated metallic behavior (dashed line). 
It has a relatively large density of electrons at zero energy and two side peaks that correspond to the Hubbard sub-bands.
Taking into account the non-local collective electronic effects via the \mbox{D-TRILEX} approach strongly suppresses the electronic density at Fermi energy (red lines).
The most remarkable change in the DOS occurs for the wide orbital, where magnetic fluctuations turn the quasiparticle peak at the Fermi energy (${E=0}$) into a pseudogap.
The density of electrons at the Fermi energy is also diminished for the narrow orbital, but this change is not that striking, because in DMFT this orbital already exhibits a pseudogap.

{\it Conclusions.}
In this work we investigated the effect of the non-local collective electronic fluctuations on the OSMT in the two-orbital Hubbard-Kanamori model with different bandwidths.
We have shown that in the absence of the interobital hopping strong magnetic fluctuations prevent the OSMT that was predicted for this model using local approximations for electronic correlations.
Instead, the system undergoes the N\'eel transition, which occurs above the Mott transition at the same critical temperature for both orbitals.  
We have shown that taking into account the interorbital hopping $t^{12}$ does not change this physical picture.
Considering $t^{12}$ only slightly shifts the N\'eel temperature upwards but does not result in a metallic
ground state that was recently predicted by DMFT~\cite{kugler2021orbitalselective}.
In this case, the N\'eel transition also occurs at higher temperatures before the OSMT is destroyed due to the presence of the inter-orbital hopping.

Our results suggest that the OSMT should rather occur in systems, where the non-local collective electronic instabilities are suppressed.
For instance, one can explore doping the system~\cite{PhysRevB.79.115119, Wang_2016, 2022arXiv220402116V}, which usually reduces the strength of magnetic fluctuations.
In this context one should avoid having van Hove singularities at Fermi energy in the electronic spectrum, because they enhance collective electronic instabilities~\cite{PhysRevB.64.165107, PhysRevB.67.125104, RevModPhys.84.299}.  
On the other hand, the dynamical orbital-selective metal-insulator transitions that are driven by collective electronic fluctuations can probably be found in systems, where the orbital degrees of freedom in these two-particle fluctuations are decoupled.
However, as we illustrated in this work, this separation of the orbital degrees of freedom cannot be provided by the Hund's exchange coupling that acts as a band decoupler for the local electronic correlations~\cite{doi:10.1146/annurev-conmatphys-020911-125045, PhysRevB.83.205112}.

\begin{acknowledgments}
The author is grateful to Alexander Lichtenstein, Silke Biermann, and Maria Chatzieleftheriou for useful discussions and comments on the work.
The author is also thankful to Matteo Vandelli and Viktor Harkov for the technical support of the \mbox{D-TRILEX} calculations.
The help of the CPHT computer support team is acknowledged as well.
The work was supported by the European Union’s Horizon 2020 Research and Innovation programme under the Marie Sk\l{}odowska Curie grant agreement No.~839551 - \mbox{2DMAGICS}.
\end{acknowledgments}

\bibliography{Ref}

\end{document}